\documentclass[12pt]{iopart}
\usepackage{color}

\usepackage{graphicx}
\newcommand{\TMYAG}{Tm$^{3+}$:YAG }
\newcommand{\NDYVO}{Nd$^{3+}$:YVO$_{4}$ }
\newcommand{\NbM}{1060 }
\newcommand{\Bw}{0.93 GHz }

\begin{document}
\title{Highly multimode memory in a crystal}
\author{M. Bonarota, J.-L. Le~Gou\"et, T. Chaneli\`ere}
\address{Laboratoire Aim\'e Cotton, CNRS-UPR 3321, Univ. Paris-Sud, B\^at. 505, 91405 Orsay cedex, France}
\ead{thierry.chaneliere@u-psud.fr}

\begin{abstract}
We experimentally demonstrate the storage of \NbM temporal modes onto a thulium-doped crystal using an atomic frequency comb (AFC). The comb covers \Bw defining the storage bandwidth. As compared to previous AFC preparation methods (pulse sequences \textit{i.e.} amplitude modulation), we only use frequency modulation to produce the desired optical pumping spectrum. To ensure an accurate spectrally selective optical pumping, the frequency-modulated laser is self-locked on the atomic comb. Our approach is general and should be applicable to a wide range of rare-earth doped material in the context of multimode quantum memory.
\end{abstract}
\pacs{42.50.Md, 42.50.Gy, 03.67.-a}
\maketitle

\section{Introduction}
The multiplexing capability of a quantum memory recently emerges as an important figure-of-merit in the prospect of long-distance quantum communication
\cite{collins2007multiplexed,SimonQrep}. The straightforward approach is to increase the number of atomic ensembles holding a quantum memory. The optically addressed mode volume compared to the total size of the medium limits the multimode capacity \cite{lan2009multiplexed}. An alternative approach consists of storing multiple temporal pulses onto a single atomic ensemble. In that perspective, the protocols exploiting the inhomogeneous broadening of rare-earth ion doped crystals (REIC) are superior \cite{nunn}. The quantum storage protocols are strongly inspired by the photon-echo technique \cite{tittel-photon} benefiting from the large ratio between inhomogeneous and homogeneous linewidth of theses materials \cite{liu2005spectroscopic}. This fine spectral resolution can be interpreted as large intrinsic time-bandwidth product or as the ability to process a train of multiple pulses in the time-domain. {This advantage is shared by the different protocols using an inhomogeneous broadening. As compared to the well established "stopped light" experiment \cite[and references therein]{OpticalQM} where the signal is mapped into the longitudinal dimension of the medium, these protocols use the excitation of spectral classes into the inhomogeneous profile. In the archetypal "Controllable Reversible Inhomogeneous Broadening" (CRIB) protocol, one initially considered the Doppler profile of atomic vapors \cite{CribGas}. It has been refined in different configurations to reach record efficiencies in both vapors \cite{Hosseini2009} and REIC \cite{HedgesNat}.}

{The recently proposed atomic frequency comb protocol (AFC) belongs to the same category \cite{AFCTh}. The prepared inhomogeneous broadening should be composed of discrete absorbing peaks defining the atomic comb.} It has the largest multimode capacity \cite{nunn}. It has been rapidly implemented in different REIC with {weak} laser pulses at the single photon level \cite{AFCsingle,AFC_NJP,Usmani2010,sabooni2009storage}. {The storage of entangled states of light has been realized very recently \cite{EntangGeneva,EntangCalgary} and represent a major breakthrough. It definitely validates the interest of this technique for quantum communication purposes. The storage mechanism can be understood by analogy with a grating. The periodic structure in the absorption profile generates an echo by diffraction in the time-domain. It has been shown to be particularly efficient with an appropriate preparation in order to create a series of narrow absorbing peaks} \cite{AFCbonarota,AFC35, sabooni2009storage}. Long storage time can be achieved by applying a Raman transfer {before the echo generation. Raman pulses actually convert back and forth the optical coherence into long-lived nuclear spin excitation freezing out the atomic evolution during the memory time \cite{AFCspin}}. It then definitely distinguishes itself from its forefather the stimulated (or three-pulse) photon-echo.

The AFC protocol requires first to prepare a spectrally periodic absorbing structure of isolated peaks. The initially smooth large inhomogeneous profile is tailored by spectral hole-burning (SHB). Multimode storage then fully exploits the spectral resolution of REIC as recently demonstrated with 64 temporal modes in \NDYVO covering a 100MHz bandwidth \cite{Usmani2010}. {The demonstration has been performed in the single photon regime and perfectly illustrates the concept of multimode storage for quantum memories}. The elementary optical pumping sequence is a pulse train composed of two \cite{AFCsingle,AFC_NJP}, three \cite{Usmani2010} or many pulses \cite{AFCbonarota}. Such an amplitude-modulation (AM) sequence exhibits a periodic optical spectrum. It is imprinted on the absorption profile because of the spectral hole-burning process. The comb bandwidth is directly given by the duration of the preparation pulses. Usmani \textit{et al.} \cite{Usmani2010} have pushed a step further this approach by applying frequency-shifted sequences to cover a wider range without being limited by the AM bandwidth. This approach could be interpreted as a mixed amplitude and frequency modulation (FM) sequence. Both AM and FM are provided by acousto-optic modulators (AOM). This technique suffers from various limitations. The AFC width will be limited by the bandwidth of the external modulator (100MHz typically for AOMs). Even if fast modulators are now available \cite{Tang:04}, they may be limited by the current electronics to produce sophisticated AM and FM sequence. Amplitude modulation is also highly demanding in terms of laser power. As the pulse duration is reduced to cover a wider band an increasing number of atoms are addressed within the inhomogeneous profile. To obtain the same population transfer with more atoms, the pulse energy should be increased accordingly. It may here be limited by the power of the continuous laser from which the pulses are cut off. Alternative solutions should be investigated.

We here propose a preparation method based on FM only. The laser frequency is directly modulated without the use of external modulators to produce a broadband optical pumping spectrum. The laser frequency should be able to continuously address a significant fraction of the inhomogeneous linewidth with a good resolution. This problem has been considered in the past for REIC in the context of optical data storage, precisely exploiting the spectro-temporal dimension of the material \cite{mitsunaga1990248, lin1995demonstration}. {Even if these demonstration cannot be extended to the single photon regime, they give a realistic estimation of the storage capacity \textit{i.e} 248 and 4000 bits for  \cite{mitsunaga1990248} and \cite{lin1995demonstration} respectively}. The agility and accuracy of chirped laser has been pushed to an unprecedented level for wideband radio-frequency analysis using REIC \cite{Crozatier:06,Gorju:07}. These achievements should be a source of inspiration for broadband and highly multimode quantum storage since they follow the same logic: laser frequency shifts can reach much larger bandwidth that the one achieved by the currently available opto-electronic devices.

The requirements and the generation of a broadband optical pumping FM spectrum will be detailed in a first section. This light will be used as a preparation beam for the protocol. We will then show that an atomic comb can be engraved over a large bandwidth in a \TMYAG crystal. We finally conclude by presenting the storage of \NbM temporal modes in the sample.

\section{Generation of a broadband FM spectrum for optical pumping}
The optical spectrum of the pumping light should produce the atomic frequency comb required for quantum storage. We will show in this section that such a spectrum can be obtained by internal modulation of an extended cavity diode laser.

\subsection{Desired spectrum}
The population dynamics relates the optical pumping spectrum to the resulting atomic comb \cite{AFCbonarota}. Sophisticated pumping scheme involving amplitude and phase modulation optimizes the efficiency of the protocol. Since we are mainly interested in highly multimode storage, we here focus on the fundamental properties of the AFC to reach this goal, namely a wide atomic comb with a large number of peaks. Intuitively, the pumping light spectrum should have the same features. The minimum spacing between the peaks depends strongly on the material properties through the optical pumping dynamics \cite{AFCsingle,AFC_NJP} (homogeneous linewidth, sources of line broadening, initial optical depth ...). A good trade-off has been found for our \TMYAG sample in the range 600-700 kHz \cite{AFC_NJP}. The spectrum width should be as large as possible, only limited by the inhomogeneous profile ($\sim$ 10 GHz). The corresponding spectrum can be produced by FM of a monochromatic laser. The modulation frequency $\nu_m$ gives the spacing between peaks. In the wideband FM situation, \textit{i.e.} the bandwidth is much larger than $\nu_m$ or equivalently the modulation index is much larger than 1, the bandwidth $B_T$ is simply two times the frequency deviation (Carson bandwidth rule). A modulation frequency $\nu_m \sim$ 600-700 kHz can be easily achieved by driving the current of a diode laser. Unfortunately a deep modulation may induce mode-hops because of the feedback from the extended cavity. We here use an extended-cavity diode laser containing an electro-optic prism (EOP) for synchronous tuning of the cavity length and the grating's incident angle \cite{Menager:00}. The fast response of the EOP can easily reach the MHz range. Large frequency deviations should be also accessible. We here propose to aim at a 1 GHz bandwidth for $B_T$ since it corresponds to the limit of our detection electronics in the pulsed regime (see section \ref{pulse}). 

\subsection{Experimental realization and observed spectra}\label{spectrum}
\begin{figure}[ht]
\centering
\includegraphics[width=14cm]{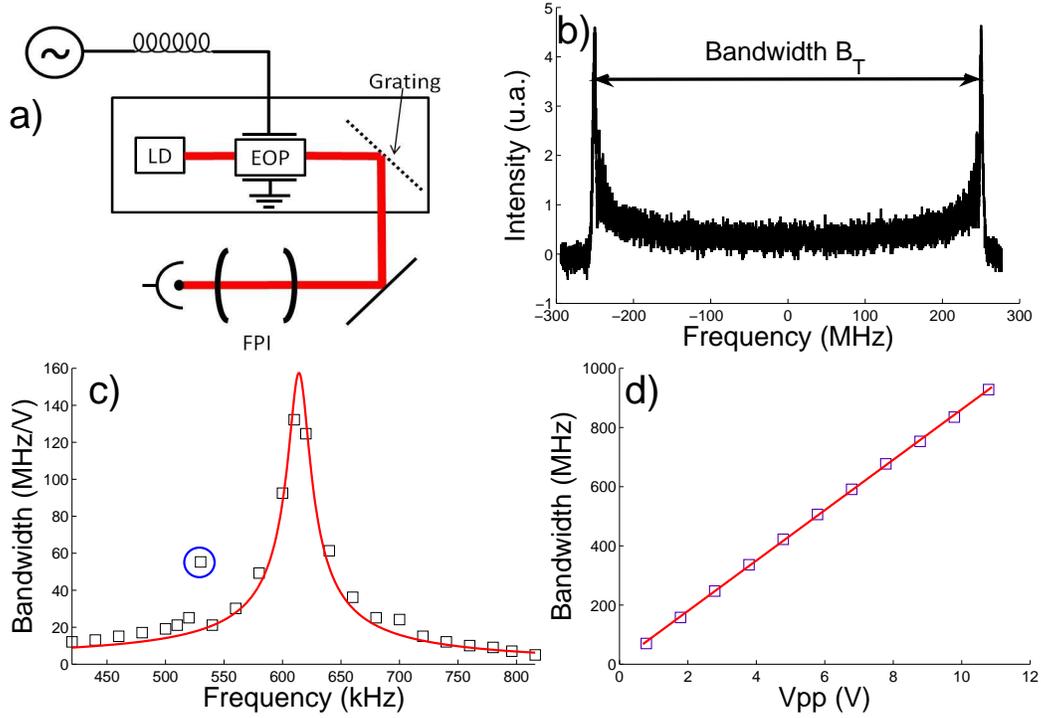}\caption{a) Extended-cavity laser diode (LD) with an intra-cavity EOP. A resonant RLC circuit enhances the moderate applied voltage. A Fabry-Perot Interferometer (FPI) is used to monitor the FM laser spectrum and measure the bandwidth $B_T$. b) Typical spectrum from the FPI. c) With a 470$\mu$ H inductor and a 55 $\Omega$ resistor, the resonance peaks at 614 kHz. The fit with the RLC circuit formula (solid red line) gives a 143 pF for the capacitor. It is consistent with the expected capacitance of the EOP electrodes (see text for details). d) Bandwidth as a function of the driving voltage. We finally step the voltage to 10.8 V$_{pp}$ corresponding to \Bw bandwidth.}
\label{fig:FIGlaser}
\end{figure}

A 1 GHz bandwidth for $B_T$ requires a large voltage driving the intra-cavity EOP. The DC response of the laser frequency to the applied voltage is typically 10 MHz/V. So to obtain $B_T=$ 1 GHz, one needs to apply  $\sim$ 100 V (the EOP response is relatively flat over a few MHz for $\nu_m$). To avoid using high-voltage amplifiers, we decide to use a resonant circuit instead. The EOP electrodes operate as the capacitor of a series RLC circuit (see figure \ref{fig:FIGlaser}a.). The resistor is $\sim$ 55 $\Omega$ including the output impedance of the generator and inductor resistance. The inductance is chosen to obtain a resonance in the 600-700 kHz range (470 $\mu$H in our case). 

To verify that the light spectrum is able to cover a large bandwidth, we use a Fabry-Perot Interferometer (FPI, Toptica FPI 100). We then directly observe in figure \ref{fig:FIGlaser}.c the FM bandwidth even if equally spaced sidebands are not resolved (few MHz resolution of the FPI). By changing the modulation frequency $\nu_m$, we measure the resonance of the RLC circuit (see figure \ref{fig:FIGlaser}.c). On this curve, we observe a small piezo-mechanical resonance of the EOP crystal at 530 kHz corresponding to previous measurements (circled marker in figure \ref{fig:FIGlaser}.c). This effect should be avoided to obtain a wideband FM spectrum. We finally conclude than close to resonance with $\nu_m =$ 626 kHz a \Bw bandwidth can be obtained for a low driving voltage (see figure \ref{fig:FIGlaser}d.). These values are set for the rest of the experiment.

The direct FM of our laser allows us to generate a spectrally periodic spectrum covering typically 1GHz. This broadband light can be used to produce the atomic comb through the optical pumping process.

\section{Atomic Frequency Comb produced by FM spectrum}
Frequency selective optical pumping or equivalently SHB is intimately related to population dynamics in the atomic system. For the optical pumping to be efficient, a long population lifetime of the shelving state is required. {There is no hyperfine structure in \TMYAG at zero magnetic field.} We here apply a field to exploit the long lifetime of the Zeeman sublevels \cite{Ohlsson:03}. It can reach few seconds at low temperature with an appropriate orientation of the field as compared to the crystalline axes \cite{louchet:035131}. The laser polarization should then be applied accordingly. We first describe the relative orientation of the magnetic field and the pumping light polarization specifically chosen. The general idea has been previously described in Refs. \cite{AFCsingle,AFC_NJP} but significant improvements have been implemented here. We will see that an active self-stabilization of the laser on the atomic comb is required as well. This experimental development is critical to obtain the long-lived atomic comb structure.

\subsection{Pumping scheme in \TMYAG}
\begin{figure}[ht]
\centering
\includegraphics[width=9cm]{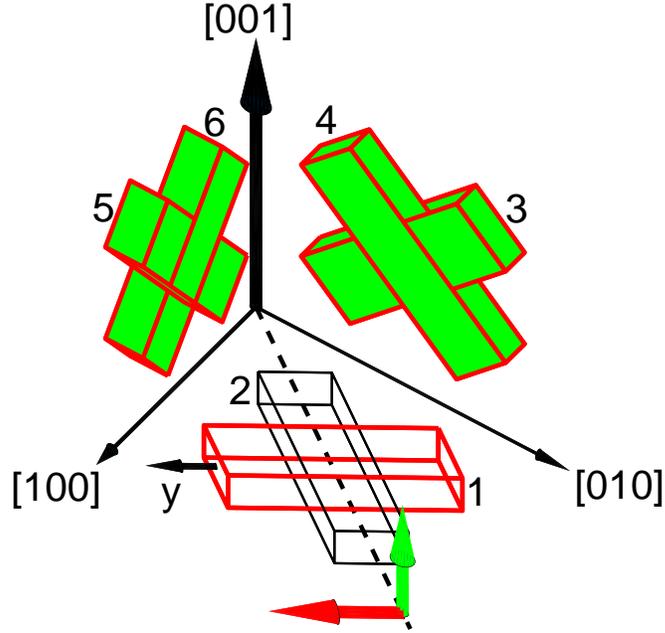}\caption{Representation of six orientationally inequivalent crystal substitution sites of thulium (1-6). The parallelepiped represents the local D2 symmetry of the sites. As an example for site 1, the transition dipole is parallel to a local axes $y$\cite{TmSites}. The laser propagates along [110] (dashed line). The polarization of preparation and the probe beam are respectively parallel to [1$\bar{\mathrm{1}}$0] (red arrow) and [001] (green arrow). We then color in red the edges (resp. in green the faces) of the sites, which are excited by the preparation (resp. probe) beam. The magnetic field is parallel to [001].}
\label{fig:SitesTmYAG}
\end{figure}

Since we use Zeeman sublevels as shelving states for the population, an accurate definition of the magnetic field, laser polarization as compared to the crystalline axes is necessary. The situation is complex in YAG because the thulium ions occupy six orientationally inequivalent crystal sites. Specific orientations of the laser polarization \cite[and references therein]{TmSites} and magnetic field \cite{louchet:035131} can exclude or render equivalent a set of different sites. To optimize the preparation of the atomic comb, the field and the polarization have been previously applied along the [001] axis of the crystal \cite{AFC_NJP} (see figure \ref{fig:SitesTmYAG}).

Even if the previous situation is globally satisfying, we here propose a refinement that may be important for weak signal storage (\textit{e.g.} single photon). The preparation procedure involves strong pumping light as compared to a weak signal to be stored. Both are well separated in time and a long delay can be inserted between them only limited by the population lifetime in the shelving state reaching 7s in our case. It is actually a major characteristic of the protocol. It allows the detection of a stored pulse at the single photon level \cite{AFCsingle,AFC_NJP}. The isolation of the probe beam requires a good extinction of the preparation beam (pump). Different techniques can be combined in that purpose. A small angle between pump and probe been can applied \cite{AFC_NJP}. Additionally acousto-optic modulators\cite{AFCsingle, AFC_NJP, sabooni2009storage} and/or a mechanical chopper \cite{AFCsingle} are used as optical switches. We here propose a configuration where pump and probe have perpendicular polarizations. Because the excitation scheme is directly related to the crystalline structure, a specific study should be done. This situation is depicted in figure \ref{fig:SitesTmYAG} for \TMYAG.

The probe beam polarization is along [001] exciting equivalently the sites 3,4,5\&6. The transition dipole is along the long dimension of the parallelepiped for each site (depicted for example as the $y$ local axes for site 1). The pump equivalently excites the same sites (with a lower  
Rabi frequency) with perpendicular polarization (parallel to [1$\bar{\mathrm{1}}$0]). It also strongly excites the site 1 which is not probed anyway. This technique should be generalized to different REIC.

After its orientation, the magnitude of the magnetic field has still to be defined. It critically influences the efficiency of the protocol \cite{AFCbonarota}. As soon as the preparation bandwidth is larger than the Zeeman splitting in the ground and the excited state, reciprocal optical pumping can destructively occur and erase the atomic comb. We cannot apply a sufficiently strong magnetic field to split the sublevels further apart because the Zeeman shift is relatively small (especially in the excited state) and our magnetic field is limited to few hundreds of gauss. A partial matching of the Zeeman splitting as compared to the comb spacing has been successfully applied \cite{AFC_NJP, AFCbonarota} to prevent reciprocal optical pumping. We here apply a magnetic field of 95G. It corresponds to the lowest magnetic field to obtain AFC storage. {Below this value, no hole-burning mechanism is observed because the lifetime of the shelving state is too short \cite{AFCbonarota}}. This low value additionally reduces the inhomogeneous broadening of the Zeeman splitting, measured to be proportional to the magnetic field amplitude.

\begin{figure}[ht]
\centering
\includegraphics[width=14cm]{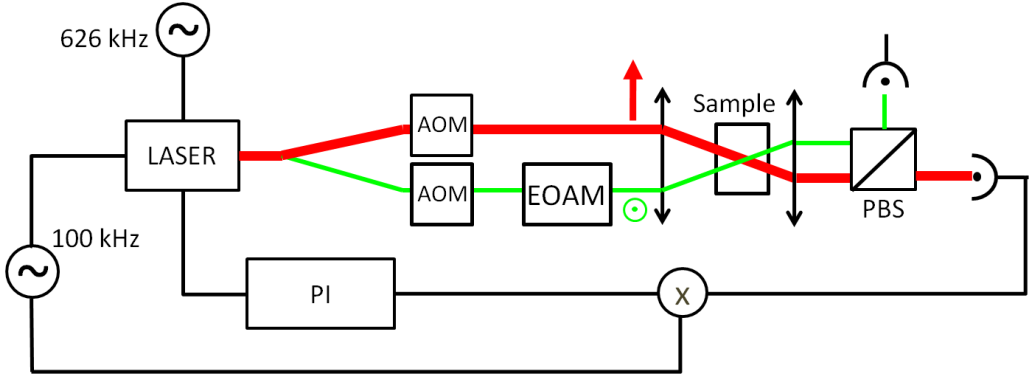}\caption{Optical setup for the preparation and probe beams. They are independently controlled by AOMs. The broadband spectrum is generated by the 626kHz modulation as described in \ref{spectrum}. A PI servo-loop is used to stabilize the pumping light on the transmission of the atomic comb (the 100 kHz modulation on the laser current is used to generate an error signal). It is extensively described in section \ref{locking} (PBS: polarizing beam splitter). An electro-optic amplitude modulator (EAOM) has been inserted on the probe beam to create ultra-short pulses to be stored (see section \ref{pulse} for details).}
\label{fig:montagetot1}
\end{figure}

We can now apply the broadband pumping light and look at the resulting comb. From the laser, we extract two beams, the pump (preparation) and the probe, independently controlled by two AOMs (see figure \ref{fig:montagetot1}). During the preparation sequence (50ms), the pumping beam is on and the laser is modulated as described in section \ref{spectrum}. We wait 5ms before probing the comb; it is sufficient to avoid the fluorescence from the excited state. During this interruption the frequency modulation is slowly switched off to prevent any perturbation of the laser. The probe beam is then monochromatic. To monitor the transmission spectrum of the atomic comb, we simply chirp the probe beam AOM frequency over few MHz. It is adequate to observe the central part of the comb (few peaks).

The previous developments are not sufficient to observe any comb-like structure. We suspect the long-term laser drift to continuously erase the structure. The effect was partially observable in real-time depending on the experimental conditions (\textit{e.g.} a smaller modulation bandwidth). The drift integrated over 7s (lifetime of the structure) should be compared to the spacing between the peaks ($\nu_n$=626 kHz in our case). This effect can be avoided by stabilizing the laser on an external reference \cite{AFC_NJP,sabooni2009storage}. It is not possible in our case because we choose to directly modulate the laser in order to reach a large bandwidth. We here implement a new locking scheme that is compatible with the FM spectrum of the laser.

\subsection{Laser self-locking on the atomic comb}\label{locking}
The preparation sequence can be interpreted as SHB by the optical pumping spectrum. SHB has been employed to externally stabilize monochromatic lasers \cite{Sellin:99,PhysRevB.62.1473,Bottger:03,Julsgaard:07,tay-2010}. So we propose to generalize this method to a broadband spectrum.

Locking on regenerative SHB is relevant because typical time response of the servo-loop matches the dynamics of the atomic systems \cite{Julsgaard:07}. This is precisely what we need. The generalization to a broadband spectrum is not obvious at first sight. The optical pumping pumping spectrum is actually composed of discrete equally spaced peaks. Each peak creates in that sense a spectral hole. A small laser drift should induce a reduction of the transmission through the SHB material because each peak goes off-resonance from its own hole. By monitoring the total transmission of the broadband light, it should be possible to compensate the laser drift and stabilize its central frequency.

The setup is described in figure \ref{fig:montagetot1}. We collect the transmission of the pumping beam. It is facilitated because pump and probe have a perpendicular polarization. To generate an error signal, we gently modulate the laser current at 100 kHz with a few percent modulation index. The error signal has the same periodic structure than the atomic comb. It is amplified, additionally integrated (PI servo-loop) and fed back to the laser current whose bandwidth is limited to 10kHz in our case. The locking scheme is complicated by the time sequence alternating pump (50ms), waiting time (5ms) and probe. The servo-loop is only active when the pump is on (longest fraction of the total time sequence). The PI parameters have been optimized by looking at the transmission spectrum during the probe sequence (atomic comb). The PI corner where proportional and integrator have the same gain is set to 300 Hz. The integrator gain is limited to 20 dB (as compare to the proportional level) at low frequency. Otherwise, the system is not stable. We attribute this behavior to electronic offsets which are integrated when the pump is off (probe sequence). As a consequence, the loop does not exhibit an integrator behavior at low frequency \cite{fox20031}. The feedback loop does not compensate extremely slow drifts. It is not an issue because the error signal is periodic over a large bandwidth (comparable to $B_T$). After a slow drift (larger than the peak spacing), the system can re-lock itself on a neighboring hole without affecting the global shape of the comb structure.

This stabilization scheme allows us to observe the atomic comb produced by the broadband pumping light.

\subsection{Resulting atomic comb}\label{comb}

\begin{figure}[ht]
\centering
\includegraphics[width=11cm]{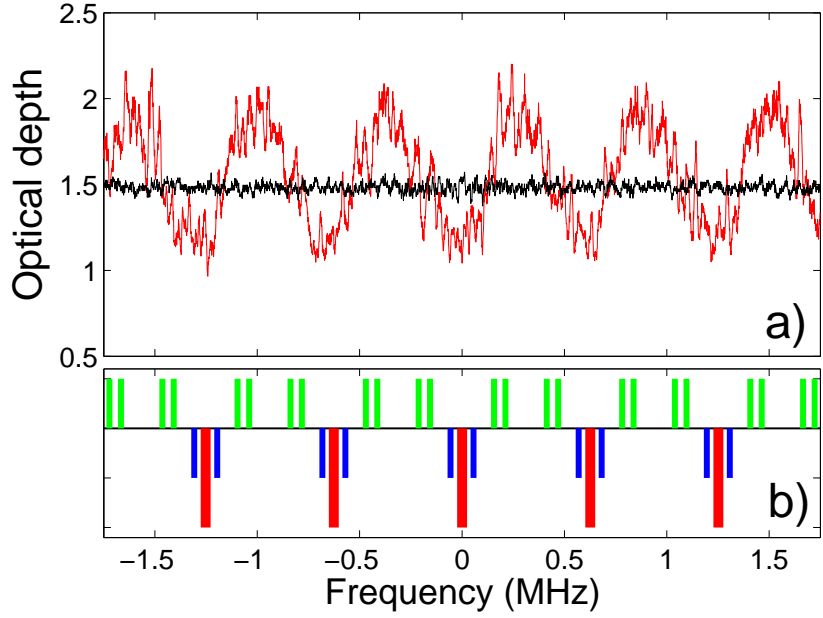}\caption{a) Observed optical depth spectra. The background (black line) is recorded when the magnetic field is off. The same flat figure is obtained without stabilization of the laser. With the self-locking technique detailed in \ref{locking}, the modulation of the pumping light is now imprinted on the atomic absorption spectrum. b) Expected positions of the holes (blue and thick red bars) and anti-holes (green bars) in \TMYAG with a magnetic field of 95G. The pumping spectrum is composed of discrete lines separated by $\nu_n$=626 kHz. This lines corresponds to the position of holes represented by thick red bars (see text for details).}
\label{fig:TraitPeigne800MHz}
\end{figure}

As described previously, the laser is rendered monochromatic by switching off the broadband modulation during the probing sequence. We here chirp the AOM to record the transmission spectra and probe the atomic populations. We then only have access to the central part of total absorption band because the AOM is scanned over few MHz only. It is sufficient to probe the atomic comb contrast by distinguishing few peaks (see figure \ref{fig:TraitPeigne800MHz}.a).

A clear comb like structure is observable. It validates our self-locking technique detailed in \ref{locking}. The contrast is limited and is {significantly} lower than previous realizations \cite{AFC_NJP, AFCbonarota}. It can be due to the imperfection of our stabilization scheme whose low-frequency gain is limited for stability reason. One can alternatively incriminate the instantaneous spectral diffusion (0.5\% Tm-doped sample \cite{liu2005spectroscopic, nilsson02}). It should now be considered because we excite a much higher bandwidth and then a much larger number of ions. This effect has still to be evaluated independantly.

The effect of reciprocal optical pumping between Zeeman sublevels should also be considered. Since the splittings in the ground $\Delta_g=2.7$ MHz and excited state $\Delta_e=0.57$ MHz are much smaller than the bandwidth, pumping and depumping occur at the same time within the inhomogeneous profile. It can equivalently be interpreted by reference to SHB spectroscopy \cite{yano}. In the present situation, a monochromatic laser would create a hole at the central frequency, two side holes positioned at $\pm \Delta_e$, four anti-holes at $\pm \Delta_g$ and $\pm \left(\Delta_g-\Delta_e\right)$ \cite{Ohlsson:03}. {Additional structures are not observed in \TMYAG because the coupling strength of the crossed optical transitions involving a spin-flip is much weaker than the direct ones (no spin-flip)} \cite{deseze}. Our broadband pumping light is actually composed of discrete lines each burning its own SHB spectrum. We have represented the expected positions of the holes and anti-holes in figure \ref{fig:TraitPeigne800MHz}.b. The blue and thick red bars mark the position of holes and the green bars the position of anti-holes. {The pumping and depumping region are ideally well separated. The intrinsic laser linewidth $\sim$ 500 kHz \cite{Gorju:07} introduces an overlap reducing the contrast of the comb. Additionally a residual slow drift of the laser can erase the comb as soon as the frequency shifts by half the comb spacing.} The interplay between these two effects certainly produces a spectrum where pumping and depumping regions are not fully separated. It may explain the moderate contrast of the resulting atomic comb. The population dynamics involving four-level systems pumped by a broadband spectrum could possibly be modeled by rate equations. This is beyond the prospect of the current proof-of-principle demonstration.

Even if we only observe the central part of the atomic comb, we expect that it is actually covering a significant fraction of the pumping light FM bandwidth $B_T$. The comb should then be able to store extremely short pulses.

\section{Multiple pulse storage}\label{pulse}

\begin{figure}[ht]
\centering
\includegraphics[width=14cm]{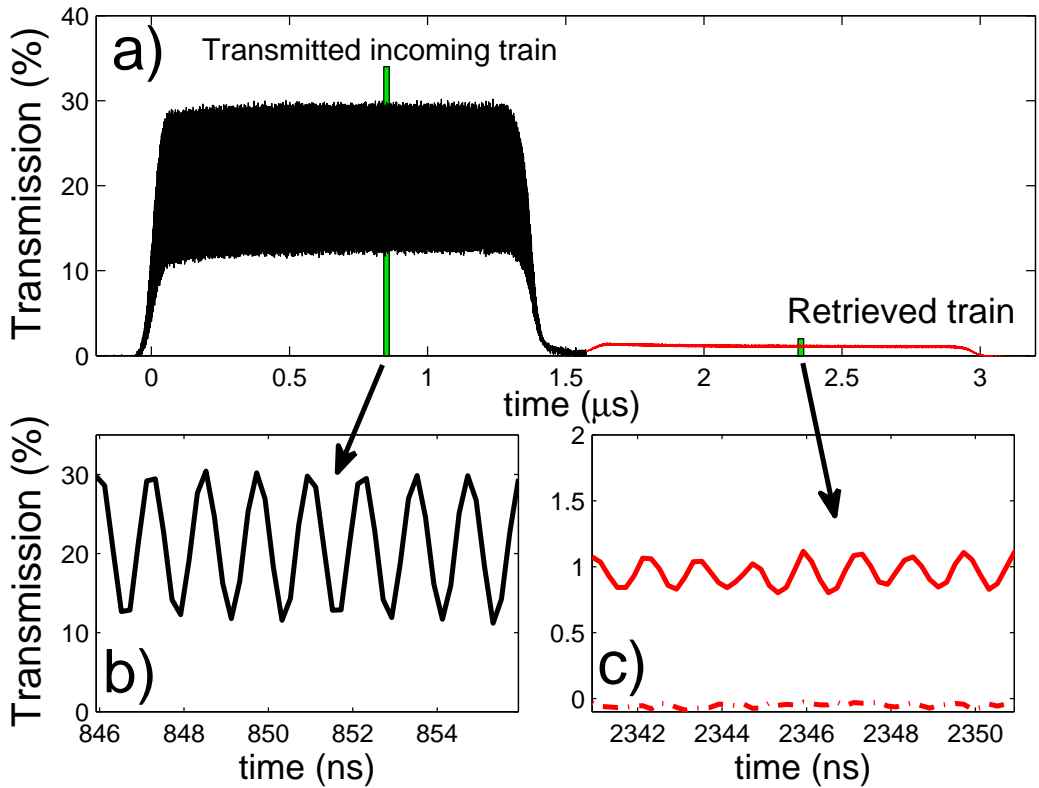}\caption{Storage of \NbM pulses. a) Incoming train transmitted by the crystal (in black), 1.6 $\mu$s later the echo appears (in red). b) The train is composed of \NbM pulses separated by 1.25 ns. c). Resolved pulses of the AFC echo. The efficiency is typically 1\% (see text for details). Since the echo is weak and to discard a potential cross-talk of the 800MHz modulation to the detection line, we switch off the magnetic field and record a reference level (no modulation is observed on the red dashed line).}
\label{fig:TraitEcho800MHzReduced_article}
\end{figure}

In order to test the storage capacity of the crystal, we sent short pulses matching the atomic comb bandwidth. This is a direct and relevant manner to verify the total comb bandwidth. A fibered Mach-Zender electro-optic modulator is inserted on the probe beam (noted EOAM in figure \ref{fig:montagetot1}. Fabricated by Alenia Marconi Systems, its bandwidth is expected to be few GHz). The main interest of the AFC protocol is its capability to store many temporal modes \textit{i.e.} a train of short pulses. The train is produced by applying a $V_\pi$ voltage modulation near the linear region of EOAM. The modulation frequency is 800 MHz chosen to be smaller than the expected comb bandwidth (\Bw). It creates a series of {identical} pulses separated by 1.25 ns (see figure \ref{fig:TraitEcho800MHzReduced_article}.b for example). {We do not have the technical possibility to generate an arbitrary sequence of pulses as required to test the full storage capacity of the memory. Nevertheless the retrieval of an identical pulse train will give us sufficient information for this future working regime}. The total duration of the train should be limited by the storage time (1.6 $\mu$s in our case, inverse of the comb peak spacing). Its duration is actually chosen to be 1.375 $\mu$s controlled by the probe beam AOM. The signal to be stored is then composed of \NbM pulses.

We clearly observe the retrieved train through the atomic comb (see figure \ref{fig:TraitEcho800MHzReduced_article}.c). The efficiency is in the 1\% range. {It is worth noting that the efficiencies are an order of magnitude lower than previously observed using the same crystal and a comparable optical thickness \cite{AFC_NJP, AFCbonarota}. As previously discussed in section \ref{comb}, the contrast of the comb is significantly lower than before because the global optical pumping dynamics is more complex (see section in \ref{comb} for a detailed discussion). To verify that the efficiency is actually limited by the poor contrast of the comb, we calculate it from the optical depth spectrum in figure \ref{fig:TraitPeigne800MHz}. We indeed apply the model presented in Refs. \cite{AFC_NJP, ChaneliereHBSM}, we expect 0.7\% agreeing relatively well the experimental result. The limiting factor is then the preparation procedure and the available optical depth of the crystal.}

Considering the absolute 800MHz modulation contrast of the incoming and retrieved train is not very significant because our detection is here limited by the electronic bandwidth of the detector (1GHz for the EOT 2030A) and the oscilloscope (1GHz for the Lecroy 104 MXi-A). {Nevertheless we clearly see that the contrast of the retrieved train is lower than the incoming one. This reduction can be attributed to a broadening of each pulse because of the limited memory bandwidth. Since we do not measure the complete transmission spectrum of the memory, we cannot really define a 3dB bandwidth. In the future working regime where each pulse should be independently controlled, a broadening effect would induce a partial overlap of the pulse and then reduce the fidelity per mode. It seems that there's a trade-off between efficiency, number of modes and actual bandwidth of the atomic comb. The exact performance of such a highly multimode memory defining the quantum communication rate has still to be evaluated.}

\section{Conclusion}
We have demonstrated the storage of \NbM temporal modes in \TMYAG using an atomic comb covering \Bw. It corresponds to a major improvement as compared to previous realizations (64 modes in \NDYVO with a 100MHz bandwidth \cite{Usmani2010}). The comb preparation technique is original because it only involves frequency-modulation. The direct modulation of the laser opens the way to very large bandwidth really exploiting the inhomogeneous broadening of REIC for quantum and classical processing applications. Even with a moderate magnetic field, we have shown that a sub-MHz spectral structure can be tailored all over the inhomogeneous profile. The experiment illustrates the broadband programming potential of closely spaced ground state sublevels. We have been able to shape the absorption profile without resorting, as usual, to distant shelving states, located outside the absorption bandwidth. In \TMYAG for instance, in applications such as the wideband spectrum analysis of an optically carried radiofrequency signal, one refreshes the spectrally periodic processing filter by continuously pumping absorbing centers into the $^{3}\textrm{F}_{4}$ bottle-neck state \cite{Lavielle:03}. This state offers little flexibility, since the lifetime is fixed at $\sim$ 10 ms. Instead, the hyperfine/Zeeman storage can be preserved for as long as $\sim$ 10 s and can be reconfigured rapidly by switching the magnetic field. Our demonstration should then be considered in a wider context.

We finally optimize the pumping scheme in our crystal (magnetic field orientation and polarization of the pumping beam). This approach should be also transposable to a wide range of rare-earth materials. Moreover, the laser is self-stabilized during the pumping sequence to ensure the engraving of the comb structure by SHB. It is a generalization of previous works for monochromatic lasers \cite{Sellin:99, PhysRevB.62.1473, Bottger:03, Julsgaard:07, tay-2010}. This stage is absolutely critical and should be considered as a general tool.

This work is supported by the European Commission through the FP7 QuRep project, by the national grant ANR-09-BLAN-0333-03 and by the Direction G\'en\'erale de l'Armement.

We thanks L. Morvan and S. Molin from Thales R\&T for providing the AMS modulator.

\section*{References}

\bibliographystyle{iopart-num}

%%%%%%%%%%%%%%%%%%%%%%%%%%%%%%%%%%%%%%%%%%%%%%%%%%%%%%%%%%
\providecommand{\newblock}{}

\end{document}